\documentclass[aps,pra,preprint,preprintnumbers,amsmath,amssymb,latexsym]{revtex4}
\usepackage{graphicx}
\usepackage{dcolumn}
\allowdisplaybreaks

\begin{document}

\title{Three-mode opto-acoustic parametric interactions with coupled cavity}

\author{H. Miao, C. Zhao, L. Ju, S. Gras, P. Barriga, Z. Zhang and D. G. Blair}
\affiliation{School of Physics, University of Western Australia, Western Australia 6009, Australia}
\date{\today}

\renewcommand{\arraystretch}{1.5}
\newcommand{\be}{\begin{equation}}
\newcommand{\ee}{\end{equation}}
\newcommand{\ba}{\begin{eqnarray}}
\newcommand{\ea}{\end{eqnarray}}

\begin{abstract}
We theoretically analyze three-mode opto-acoustic parametric interactions in a coupled Fabry-Perot
cavity, where one acoustic mode interacts with two optical modes. We show explicitly that extra
degrees of freedom in a coupled cavity allow explorations of both parametric instability and cooling
regimes with high parametric gain in a single table top experiment. This work can motivate experimental
realizations of the three-mode parametric instability which might be an issue in next-generation
gravitational-wave detectors with high optical-power cavities, helping in the development of better
models, and in developing techniques for controls. In addition, we show that
the same scheme can be implemented in the resolved-sideband acoustic-mode cooling.
\end{abstract}


\maketitle

\section{Introduction}

Parametric interactions have wide applications and arouse great interests in various fields of
physics, from high-sensitivity transducers to low-noise amplifiers and optical parametric oscillators.
Two-mode parametric interactions have been used to cool acoustic modes of mechanical oscillators. In
Ref. \cite{Blair}, the authors cooled the normal mode of a 1.5 tonne Nb bar down to 5mK by using a microwave
parametric transducer. With the same principle, a high-frequency acoustic mode of a nano-mechanical
oscillator is cooled near its quantum ground state through coupling to a superconducting single-electron
transistor \cite{Naik}. More recently,  by coupling mechanical resonators to optical cavities, various
table top experiments have demonstrated significant cooling of acoustic modes \cite{Gigan,Arcizet,Kleckner,Schliesser1,Corbitt1,Corbitt2,Schliesser2,Poggio,Thompson,Lowry,Schediwy}.
These experiments show great potential of achieving the quantum ground state of macroscopic mechanical
oscillators, which would be a breakthrough in both theoretical and experimental physics.

Three-mode opto-acoustic parametric interactions were first investigated by Braginsky {\it et al.}
\cite{braginsky,braginskyII} in the context of high optical-power Fabry-Perot cavities for interferometric
gravitational-wave detectors. It was shown that three-mode interactions led to a risk of parametric
instability (PI) in which the amplitude of acoustic modes of test masses could grow exponentially,
thus undermining sensitivities of the detectors. The cause of parametric instability is the radiation
pressure mediated coupling between optical modes and acoustic modes. It can occur if the shape of
high-order cavity modes, hereafter denoted by TEM$_{mn}$, have a substantial overlap with that of
acoustic modes and simultaneously if the mode gap between TEM$_{mn}$ and TEM$_{00}$ is equal to acoustic-mode frequency
up to an error of the linewidth of cavity modes. Further theoretical studies have followed up this
pioneering work. Zhao {\it et al.} \cite{zhao} extended their analysis and took into account 3D structures
of the optical and acoustic modes. Ju {\it et al.} \cite{Ju} further considered the overall contributions
from multiple optical modes, showing that multiple interactions increase the risk of instability.
Gurkovsky {\it et al.} \cite{Gurkovsky} analyzed PI in a signal-recycled (SR) interferometer and shows
that the chance of PI was reduced due to the narrow linewidth of SR interferometer. Additionally, many
ideas have been proposed to prevent PI, e.g., changing the radius of curvature (RoC) of the mirror
\cite{zhao}, using an optical spring tranquilizer \cite{braginskyIII} or a ring damper \cite{Gras}.
In the light of the above predictions, it is important to develop experimental techniques for their
investigations. In recent progress, the University of Western Australia group demonstrated three-mode
interactions in an 80m-long Fabry-Perot cavity \cite{AIGO}. They used a compensation plate as a thermal
lens to tune the effective RoC of the mirror, which changes the Gouy phase or equivalently the mode gap,
such that the resonant condition mentioned is satisfied. By capacitively exciting an acoustic mode to
interact with the $\rm TEM_{00}$ mode, they observed resonance of a high-order optical mode at the predicted
frequency. This experiment confirmed the principle of three-mode interactions, but due to low optical
power and small overlap, it had not yet achieved self-sustained parametric instability or cooling.
In this paper, we propose an alternative method of tuning the Gouy phase. It was first introduced by Mueller
\cite{Guido} as a means of designing a stable recycling cavity for next-generation gravitational-wave
detectors. By adding a mirror and a lens onto a single cavity, which forms a coupled cavity, one can
obtain any desired Gouy phase shift simply by adjusting relative positions of these optical components.
In following sections, we will show explicitly that a table top experiment can be set up, allowing
observations of three-mode interactions with high parametric gain. This can help us better understand
possible parametric instability in interferometric gravitational-wave detectors, and can also be applied, in general, as a
design concept for opto-acoustic parametric amplifiers, specifically to the resolved-sideband cooling
\cite{Marquart, Rae} of acoustic modes.

This paper is organized as follows: In Sec. II, we will review the theory of three-mode opto-acoustic
interactions. In Sec. III, we will analyze three-mode interactions in a coupled cavity. Finally,
we will summarize our results in Sec IV.

\section{Review of three-mode opto-acoustic parametric interaction theory}

For the coherence of this article, we will briefly review the theory of three-mode opto-acoustic parametric
interactions. They can be understood qualitatively in both quantum and classical pictures. In the former
picture, we can treat the Fabry-Perot cavity as a multi-level quantum system occupied by photons, while the
mechanical oscillator (end mirror) consists of phonon (acoustic) modes in different energy eigenstates. For
simplicity, we only consider one phonon mode, which represents the situation in this proposed scheme.
\begin{figure}
\includegraphics[width=11cm,bb=71 315 483 491,clip]{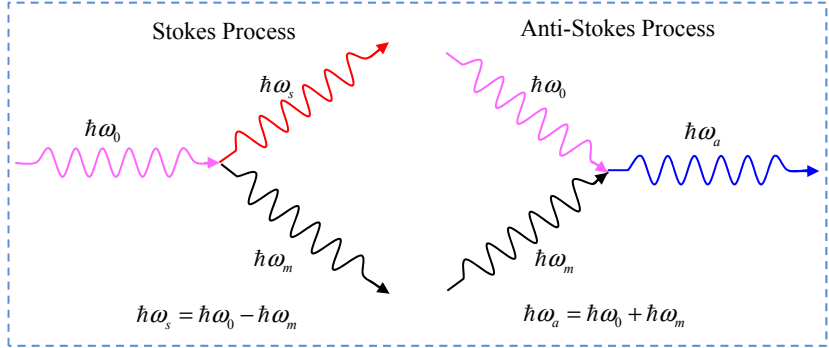}\caption{
\label{interaction} Parametric interactions between photons and phonons. In the Stokes process,
the photon is down converted into a photon with lower energy and at the same time one phonon
is created. This process will transfer energy from the optical field into the mechanical degree of
freedom, thus amplifying the acoustic mode. In the Anti-Stokes process,
the photon is scattered into a higher-frequency photon, accompanied by annihilation of one
phonon. In this process, the mechanical oscillation is damped. Classically, these two processes simply correspond to
modulations of light by mechanical oscillations into two sidebands at frequencies $\omega_0\pm\omega_m$.}
\end{figure}
Initially, the photons stay in one eigenstate (the TEM$_{00}$ mode). Resonant amplification occurs when
the photons are scattered by the phonons into another eignstate (the TEM$_{mn}$ mode). The phonons can
absorb energy from the photons (conventionally known as a Stokes process in Raman scattering), causing instability.
Alternatively, they can release energy into the optical field (Anti-Stokes process), which is also called cooling
in the literature. We show both processes schematically in Fig. \ref{interaction}. In the classical
picture \cite{braginsky}, mechanical oscillations modulate the carrier frequency $\omega_0$ into two sidebands.
The lower sideband corresponds to the Stokes mode with frequency $\omega_s=\omega_0-\omega_m$ and
the upper sideband is the Anti-Stokes mode with $\omega_a=\omega_0+\omega_m$. If they match the frequency
of TEM$_{mn}$ $\omega_1$, higher-order optical mode will be excited. In turn, these resonant optical modes
TEM$_{mn}$ and TEM$_{00}$ will exert a radiation pressure on the mechanical oscillator at the beating frequency $|\omega_1-\omega_0|=\omega_m$.
Depending on the phase, this force will either amplify or damp the motion of the oscillator. To achieve
good coupling, a substantial spatial overlap between the acoustic mode and the TEM$_{mn}$ mode is required.

To quantify the interactions, we follow the formalism in Ref. \cite{braginsky}. In the paper, they introduced
parametric gain $\mathcal{R}$ to quantify the strength of three-mode interactions, which is defined as,
\be \label{R} {\mathcal{R}}=\frac{4I_c Q_m}{m
cL\omega_{m}^{2}}
\left[\frac{Q_s\Lambda_s}{1+(\Delta\omega_s/\gamma_s)^2}-
\frac{Q_{a}\Lambda_{a}}{1+(\Delta\omega_{a}/\gamma_a)^2}\right].
\ee
Here subscript $s$ denotes Stokes mode and $a$ for Anti-Stokes mode; $L$ is the total length of the cavity;
$I_c$ is the intra-cavity power; $m$ is the mass and $Q_m$ is the mechanical quality factor. The detunings $\Delta\omega_{s(a)}
\equiv|\omega_{s(a)}-\omega_1|$ and $\gamma_{a(s)}$ is decay rates of optical modes. Overlapping factor $\Lambda_{a(s)}$
is given by,
\be \Lambda_{a(s)}=\frac{V(\int
f_0(\vec{r}_{\perp})f_{a(s)}(\vec{r}_{\perp})u_zd\vec{r}_{\perp})^2}
{\int |f_0|^2d\vec{r}_{\perp}\int |f_{a(s)}|^2d\vec{r}_{\perp}\int
|\vec{u}|^2dV}, \ee
where $f_{0,a,s}$ is optical-mode shapes; $u_z$ is the component of $\vec{u}$ normal to the mirror surface;
The integrals $\int d\vec{r}_{\perp}$ and $\int dV$ correspond to integration over the mirror surface and volume
$V$ respectively. For simplicity, we assume that $\Lambda_{s(a)}\sim 1$, which can be achieved in real
experiments. In the tuned case that $\Delta\omega_{s(a)}=0$, $\mathcal{R}$ can be simply written as
\be\label{gains}
{\mathcal{R}}=\pm\frac{4I_c Q_m Q_{s(a)}}{mcL\omega_{m}^{2}},
\ee
with $+$ for the Stokes process and $-$ for the Anti-Stokes process. The resulting decay rate of
the acoustic mode $\gamma_m'$ can be written as
\be\label{dm}
\gamma_{m}'=\frac{1}{2}\left[(\gamma_{s(a)}+\gamma_m)-
\sqrt{(\gamma_{s(a)}-\gamma_m)^2+4{\mathcal{R}}\gamma_{s(a)}\gamma_m}\right].
\ee
Since usually $\gamma_{a(s)}\gg \gamma_m$, then
\be
\gamma_{m}'\approx(1-{\mathcal{R}})\gamma_m.
\ee
When ${\mathcal{R}}=0$, we obtain the trivial case $\gamma_m'=\gamma_m$. For the Stokes process with ${\mathcal{R}}>0$
(positive gain), the decay rate of the acoustic mode decreases and the mechanical oscillation will
be amplified. PI happens if ${\cal R}>1$, where we have $\gamma_m'<0$. For the Anti-Stokes process with
${\mathcal{R}}<0$ (negative gain), the acoustic mode will be cooled as the effective decay rate increases.

\section{Three-mode interactions with a coupled cavity}

\begin{figure}[!h]
\includegraphics[width=8.1cm, bb=67 284 394 412 clip]{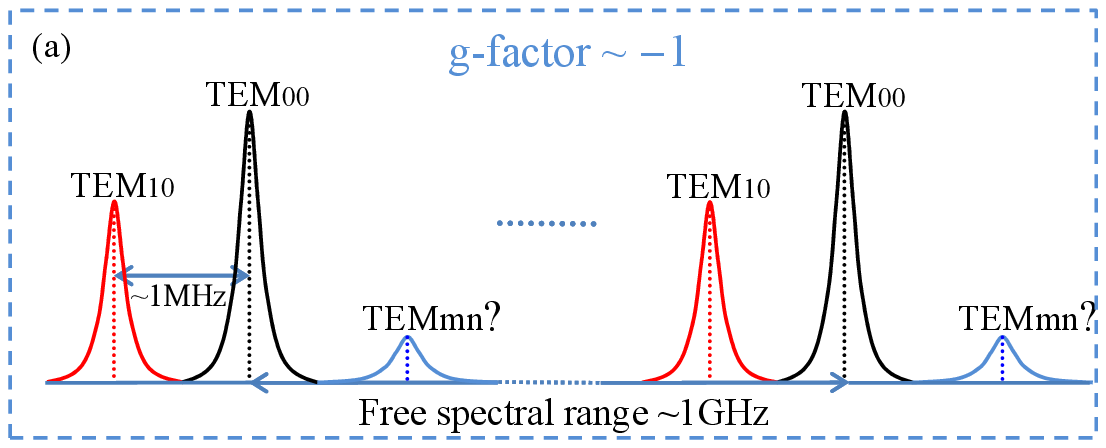}
\includegraphics[width=8.1cm, bb=72 299 400 426,clip]{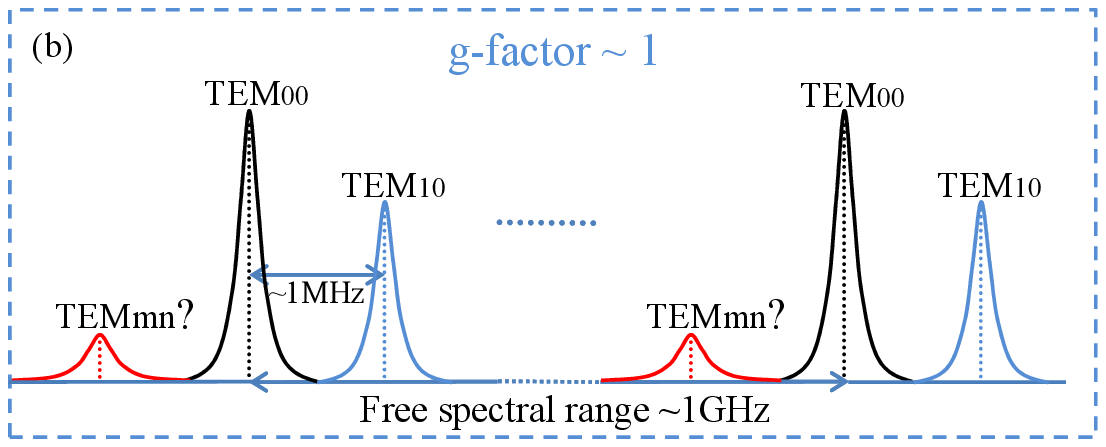}
\caption{
\label{modes} The optical modes of single Fabry-Perot cavity with length
$\sim 10$ cm. The panel (a) shows the mode distribution for a near-concentric cavity with g-factor
$\sim -1$ which suits for observing PI,  while panel (b) is the near-planar case with g-factor
$\sim 1$ which suits for cooling experiment. In both cases, there are no symmetric modes on the opposite side of
the TEM$_{00}$ mode because the higher order mode TEM$_{mn}$ marked with `?' are highly lossy due to
diffraction losses. This is preferred for experimental realizations of three-mode interactions because
we know from Eq. \eqref{R} that any symmetric mode on the opposite side of the TEM$_{00}$ mode will reduce the
absolute value of parametric gain. However, both cavities are marginally
stable and very susceptible to misalignment.}
\end{figure}

In this section, we will discuss how to explore three-mode interactions using a coupled cavity. To make
our analysis close to realistic experiments, we consider a torsional acoustic mode with frequency $\sim 1$
MHz interacting with the optical TEM$_{10}$ and TEM$_{00}$ modes. The configuration is chosen because MHz
can be easily achieved in a mm-scale structure and the torsional mode has a large spatial overlap with the
TEM$_{10}$ mode.

\begin{figure}
\includegraphics[width=10cm, bb=112 359 500 496,clip]{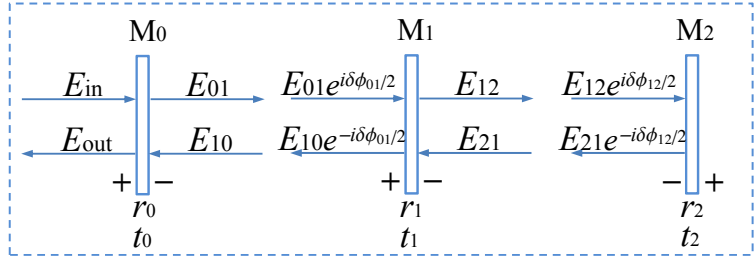}\caption{The optical fields of
the coupled cavity. Here $\delta\phi_{01, 12}$ are the round-trip phase shift of light in the sub-cavity
(formed by M$_0$ and M$_1$) and main cavity (formed by M$_1$ and M$_2$) respectively. We use the convention
that the mirrors have minus reflectivity on the side with a coating layer.}
\label{fields}
\end{figure}

To begin with, let us consider a single Fabry-Perot cavity to see why a coupled cavity is necessary.
The free spectral range of a single cavity with length $\sim 10$ cm is approximately 1GHz. Therefore, as
shown in Fig. \ref{modes}, one has to build either a near-planar or near-concentric cavity to obtain
a desired mode gap around 1MHz between TEM$_{10}$ and TEM$_{00}$. For both cases, the cavity is marginally
stable and susceptible to misalignment. It is also difficult to get accesses to both instability and cooling
regimes in a single setup. Creating a coupled Fabry-Perot cavity solves these problems. As we will show later, the resulting scheme is
stable. Additionally, we can easily tune between instability and cooling regimes. The coupled cavity is showed
schematically in Fig. \ref{fields}. It is similar to the configuration of power or signal recycling interferometers \cite{Meers, chen}
when one considers either the common mode or the differential mode. The field dynamics can be easily obtained as shown
in Ref. \cite{Rakhmanov} by treating the sub-cavity as an effective mirror with frequency and mode-dependent
transmissivity and reflectivity. Specifically, the effective transmissivity $t_{01}$ is

\begin{figure}
\centerline{\includegraphics[width=10cm, bb=143 476 518 570,clip]{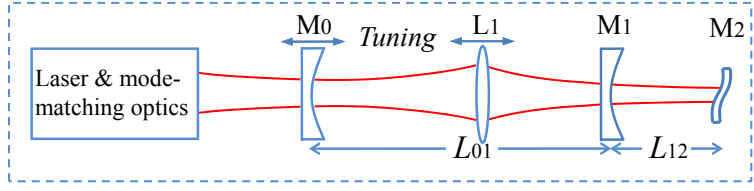}}\caption{ \label{optical layout}
The optical layout for the table top experiment, where a 1 MHz micro torsional oscillator (M$_2$) interacts with
the optical TEM$_{10}$ mode and the TEM$_{00}$ mode. By tuning the positions of mirror M$_0$ and lens L$_1$, we
can continuously change the frequency of TEM$_{10}$. If the losses in L$_1$ were an issue, it could easily be
replaced by a concave mirror.}
\end{figure}

\be\label{t01}
t_{01}\equiv\frac{E_{12}}{E_{\rm in}}=\frac{t_0t_1}{1+r_0r_1e^{i\delta\phi_{01}}},
\ee
and the effective reflectivity $r_{10}$ is given by
\be
r_{10}\equiv\frac{E_{12}}{E_{21}}=-r_1-\frac{t_1^2r_0e^{i\delta\phi_{01}}}{1+r_0r_1e^{i\delta\phi_{01}}}.
\ee
The corresponding $E_{12}$ inside the main cavity can be written as
\be \label{E12}
E_{12}=\frac{E_{\rm in}t_{01}}{1+r_{10}r_2e^{i\delta\phi_{12}}}
=\frac{E_{\rm in}t_{01}}{1-|r_{10}|r_2e^{i[\arg(r_{10})+\delta\phi_{12}+\pi]}}.
\ee
The resonance occurs when the phase factor in Eq. \eqref{E12} is equal to $2n\pi$, which critically depends
upon the phase angle of the effective reflectivity, namely $\arg(r_{10})$. Specifically, when the TEM$_{00}$
mode resonates inside the main cavity, which requires that $\delta\phi_{01}^{\rm TEM_{00}}=\delta\phi_{12}
^{\rm TEM_{00}}=2n\pi$, the phase shift of the TEM$_{10}$ mode $\delta\phi_{ij}^{\rm TEM_{10}}$ is
\be \label{dfHOM}
\delta\phi_{ij}^{\rm TEM_{10}}=\frac{2
L_{ij}}{c}\Delta\omega-2\Phi_g^{ij}+2n'\pi,~~ij=01,12,
\ee
where $\Delta\omega\equiv\omega_1-\omega_0$ is the mode gap between TEM$_{10}$ and TEM$_{00}$; $\Phi_g$
is the Gouy phase and $n, n'$ are integers. In order to satisfy the resonant condition for three-mode
interactions, we need to adjust $\delta\phi_{01}^{\rm TEM_{10}}$, which changes $\arg(r_{10})$, such
that $\Delta\omega=\pm \omega_m$. To achieve this, one obvious way is to change the length of sub-cavity
$L_{01}$, but this turns out to be impractical due to a small tuning range. An alternative and more
practical approach, as shown in Fig. \ref{optical layout}, is to add another lens or concave mirror inside
the sub-cavity to tune the Gouy phase $\Phi_g^{01}$. The resulting scheme is similar to the proposed stable
recycling cavity for next-generation gravitational-wave detectors \cite{Guido}. With the additional lens, the
Gaussian beam gets focused inside the sub-cavity. Since the Gouy phase changes almost from $-\frac{\pi}{2}$ to
$\frac{\pi}{2}$ within one Rayleigh range around the waist, one can easily obtain a desired $\delta\phi_{01}
^{\rm TEM_{10}}$ simply by adjusting the position of M$_0$ near the waist. This might lead to problems with
power density due to the small waist size, but for the table top experiment we consider here, the power
density is quite low.

The corresponding $\Phi_g^{01}$ with an additional lens can be derived straightforwardly by using
the ray transfer relation for a Gaussian beam, which is given by
\be
q'=\frac{f q}{f-q}.
\ee
Here $f$ is the focal length of $\rm L_1$; $q^{(')}\equiv z^{(')}+iz_{R}^{(')}$; $z$ is the displacement
relative to the waist; $z_R$ is the Rayleigh range and superscript $'$ denotes quantities after the lens.
This dictates
\ba \label{znew} z'&=&\frac{f(z
f-z^2-z_{R}^2)}{(f-z)^2+z_{R}^2},\\
z_{R}'&=&\frac{z_{R}f^2}{(f-z)^2+z_{R}^2}.
\ea
The resulting Gouy phase at any point is given by
\be
\label{gp}
\Phi_g(z)=\Bigg\{
\begin{array}{ll}
\arctan\left(z/z_R\right),&z<z_L\\
\arctan\left[(z-z_L+z_L')/z_R'\right]+\arctan\left(z_L/z_R\right)-\arctan\left(z_L'/z_R'\right),&z\geq
z_L
\end{array}
\ee
where $z_L^{(')}$ is the position of $\rm L_1$ relative to the waist. Gouy phase $\Phi_g^{01}$ is the
difference between wavefront at $\rm M_0$ and $\rm M_1$, namely
\be
\Phi_g^{01}=\Phi_g(z_{\rm M_0})-\Phi_g(z_{\rm M_1}),
\label{15}
\ee
where $z_{\rm M_0}$ and $z_{\rm M_1}$ are the positions of $\rm M_0$ and $\rm M_1$ relative to the waist
respectively. Therefore, by adjusting the positions of $\rm M_0$ and $\rm L_1$ as shown in Fig. \ref{optical layout},
we can continuously tune $\Phi_g^{01}$ such that $\Delta\omega=\omega_m$.

Equations \eqref{t01} to \eqref{15} provide the design tools of the coupled cavity for three-mode interactions.
To realize the experiment, we first need to design the main cavity and specify $L_{12}$, $\omega_m$, the
RoCs of $\rm M_0, M_1, M_2$ and the focal length of $\rm L_1$. From Eq. \eqref{E12} and Eq. \eqref{dfHOM},
we can find out the required $\arg(r_{01})$ which gives the right mode gap between TEM$_{10}$ and $\rm TEM_{00}$.
This will gives us one constraint. Combining with the requirement of mode matching to
$\rm M_0$, we can fix two degrees of freedom of the system, namely the positions of $\rm M_0$ and $\rm L_1$.
To demonstrate this principle explicitly, we present a solution that is close to a realistic experimental setup.
We assume the following,
\[
\begin{array}{lll}
\mbox{L}_{12}=75~\mbox{mm}&\omega_m=1\mbox{MHz}&f=100~\mbox{mm}\\
R_0=500~\mbox{mm}&r_0=\sqrt{0.999}&A_0=500~\mbox{ppm}\\
R_1=100~\mbox{mm}&r_1=\sqrt{0.9}&A_1=500~\mbox{ppm}\\
R_2=\infty~\mbox{mm}&r_2=\sqrt{0.9995}&A_2=500~\mbox{ppm}.
\end{array}
\]
Here $R_{i} (i=0,1,2)$ are RoCs; $r_i$ denotes the amplitude reflectivity; $t_i$ is the amplitude
transmissivity and $A_i$ is the optical loss which satisfy $r_i^2+t_i^2+A_i=1~(i=0,1,2)$.

\begin{figure}
\includegraphics[width=8cm,bb=79 317 365 412,clip]{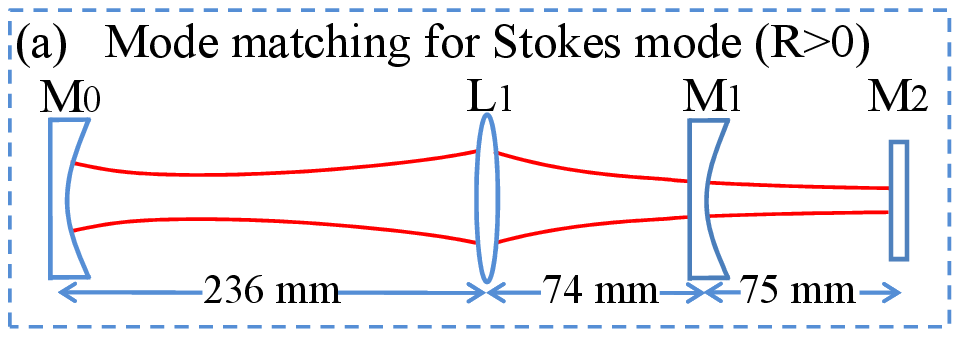}
\includegraphics[width=8cm,bb=69 317 355 412,clip]{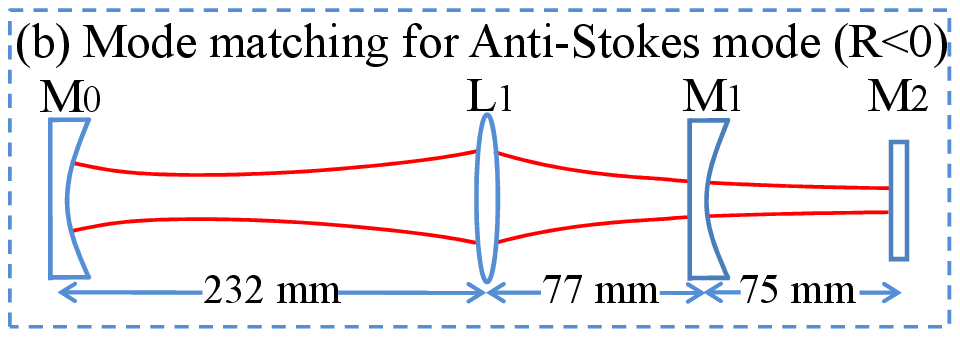}\caption{
\label{modematching} The mode matching for the positive gain and
negative gain by adjusting the relative position of $\rm M_0$
and $\rm L_1$. Only small adjustment is needed to tune from
one case to another.}
\end{figure}
\begin{figure}
\includegraphics[width=8cm,bb=25 9 327 230,clip]{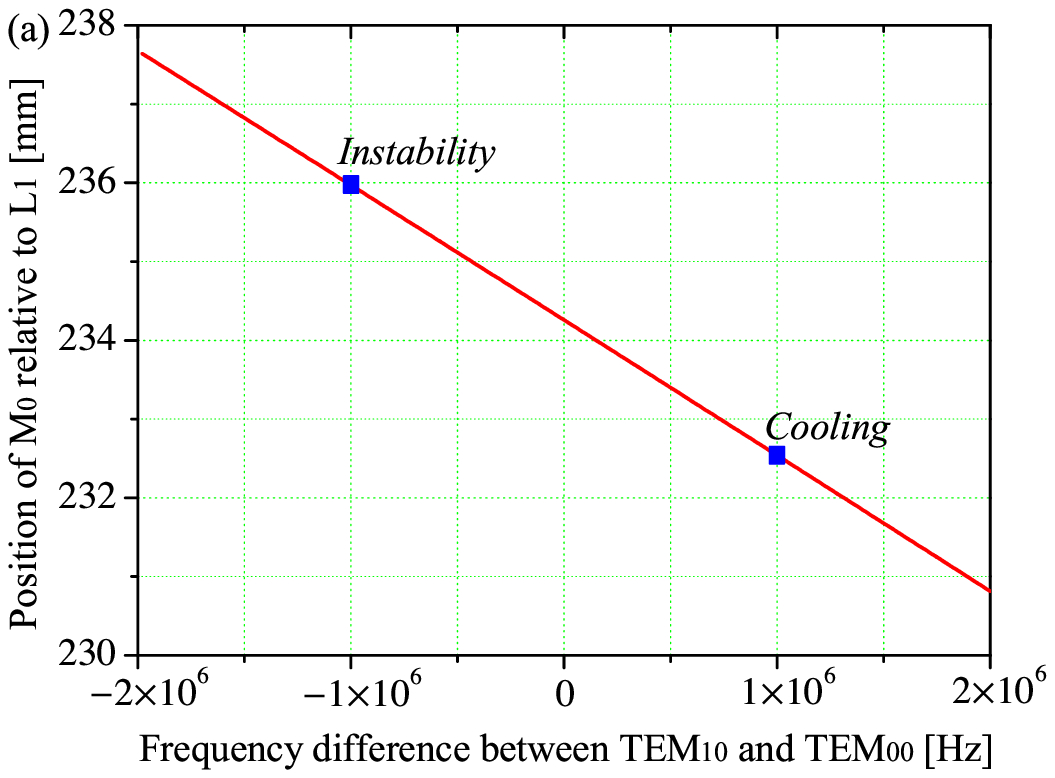}
\includegraphics[width=8cm,bb=29 9 331 228,clip]{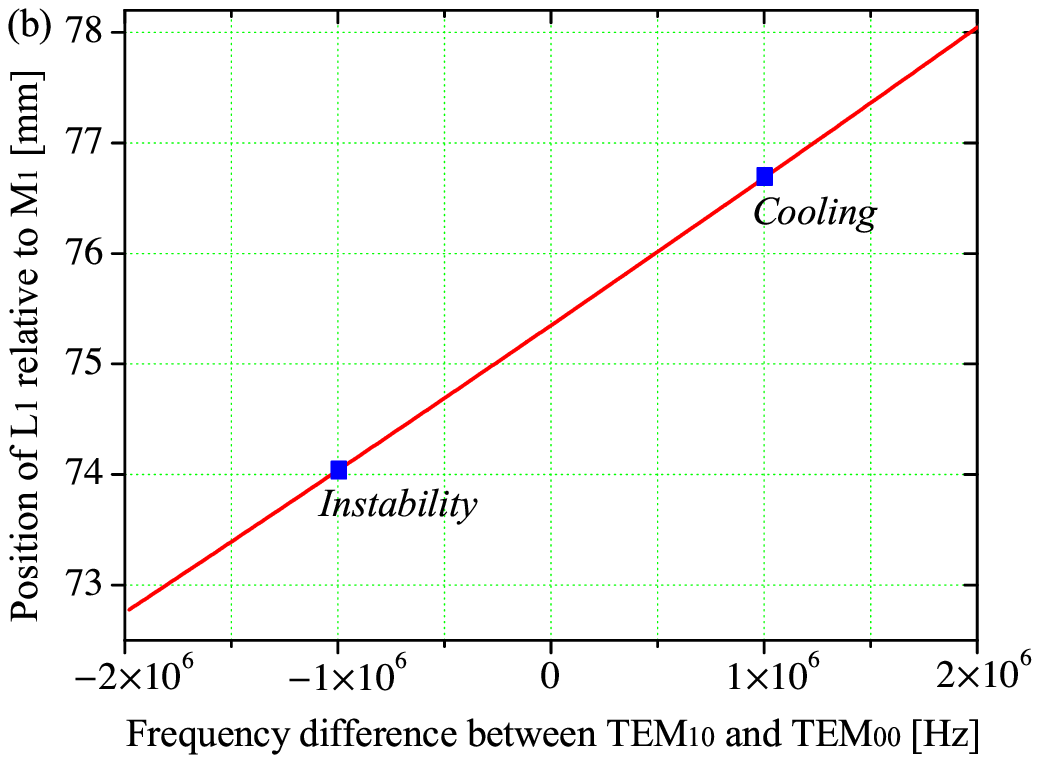}
\caption{
\label{tuning} Panels (a) and (b) show the mode gap between TEM$_{10}$ and TEM$_{00}$ as
a function of position of M$_0$ relative to L$_1$ and position of L$_1$ relative to M$_1$ respectively.
The dots in both figures are the situations considered in Fig. \ref{modematching}. Clearly, we can
tune between the instability and cooling regimes continuously.}
\end{figure}
\begin{figure}
\includegraphics[width=8cm,bb=25 13 328 226,clip]{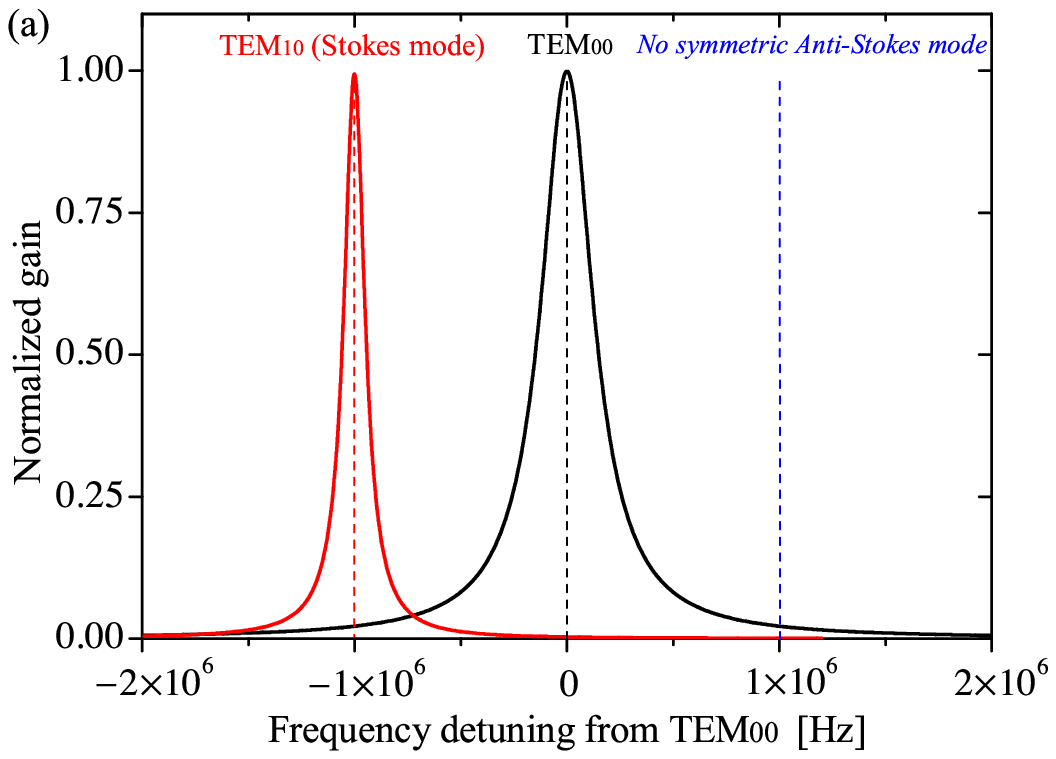}
\includegraphics[width=8cm,bb=25 10 325 228,clip]{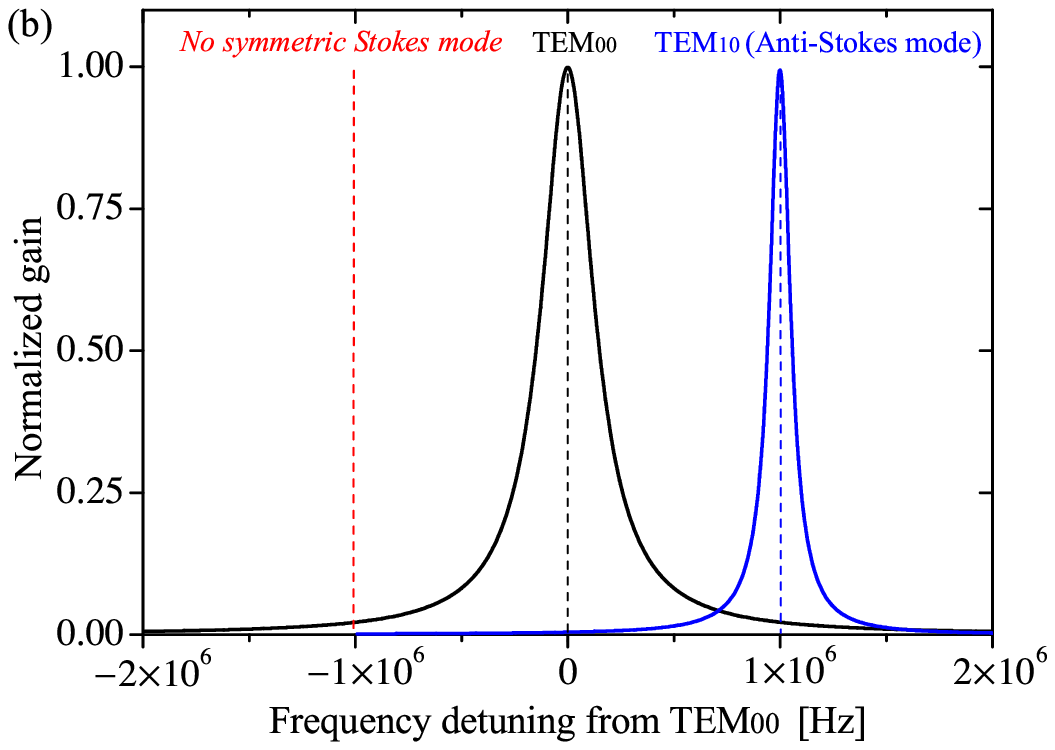}
\caption{ \label{gain} The normalized gain of the TEM$_{00}$ mode and the TEM$_{10}$ mode in the positive
and negative gain configurations. The mode gap is equal to $\omega_m\sim$ 1MHz, which fulfils the resonant
condition for the three-mode opto-acoustic interactions. Here we simply assume that the size of the mirrors
is infinite so the quality factor of the TEM$_{10}$ mode is solely due to optical losses such as absorbtion.
This assumption is reasonable when the mode number is small. Given the specifications in the main text,
$Q_a\approx Q_s=2.4\times 10^9$ and $\omega_m/\gamma_{a} <1$. Therefore, it can be implemented in the
resolved-sideband cooling. (a) The TEM$_{10}$ mode is 1 MHz below the TEM$_{00}$
mode; (b)The TEM$_{10}$ mode is 1 MHz above the TEM$_{00}$ mode.}
\end{figure}

The results of mode matching for both positive (instability) and negative gain (cooling) configurations are
shown in Fig. \ref{modematching}. In Fig. \ref{tuning}, we show the mode gap between TEM$_{10}$ and TEM$_{00}$
as a function of the position of M$_{0}$ relative to L$_1$ and the position of L$_1$ relative to M$_1$. In this
particular case, the dependence is almost linear with a slope $\sim 2\mbox{mm}/\mbox{MHz}$ for both panels. This
indicates that to tune within a cavity linewidth $\sim 0.1$ MHz, it requires the mirror position to be adjusted
within several $100\mu{\rm m}$, which can be achieved easily. Therefore, we can continuously tune between
instability and cooling regimes. Fig. \ref{gain} shows the resulting resonance curves for both cases with
the corresponding mode matching shown in Fig. \ref{modematching}. The corresponding mode gap between the TEM$_{10}$ and
TEM$_{00}$ modes is equal to $\omega_m\sim {\rm 1MHz}$. More importantly, there is no symmetric mode
on the opposite side of the TEM$_{00}$ mode, whose presence could contribute a parametric gain with the opposite
sign, thereby suppressing the overall effects. The absolute value of parametric gain ${\mathcal{R}}$ could be larger than 1, if we
further assume that the intra-cavity power $I_c$ is 100 mW, $Q_m=10^6$, the mass of the oscillator $m=1$ mg and the wavelength
of light is 1064 nm. Since the cavity is in the resolved-sideband regime where the cavity linewidth is much smaller than
the mechanical frequency  \cite{Marquart}, this configuration can also be applied in the resolved-sideband cooling
of acoustic modes, which is less susceptible to quantum noise. Although the quantum noise analysis in Refs.
\cite{Marquart, Rae, Genes} only discusses a two-mode system, their results can be extended to the three-mode system
by viewing the $\rm TEM_{00}$ mode as the carrier light. This can be justified by writing down the Hamiltonian
of this system, which is
\be
\hat{H}=\frac{\hat{p}^2}{2m}+\frac{1}{2}m\omega_m^2\hat{x}^2+\hbar\omega_0\hat{a}^{\dag}\hat{a}+\hbar\omega_1\hat{b}^{\dag}\hat{b}+
\hbar G_0 \hat{x}(\hat{a}^{\dag}\hat{b}+\hat{b}^{\dag}\hat{a})+\hat{H}_{\rm drive},
\ee
Here the first two terms represent the energy of the acoustic mode, where $\hat{x}$ is displacement and $\hat{p}$ is
momentum which satisfy the commutation relation $[\hat{x},\hat{p}]=i\hbar$; $\hat{a}$ and $\hat{b}$ are the
annihilation operators of the $\rm TEM_{00}$ and $\rm TEM_{10}$ modes respectively; the fifth term quantifies
the interactions between the optical modes and the acoustic mode with the coupling constant $G_0\equiv \Lambda_{a, s}\omega_0/L$;
the last term corresponds to external driving. Since only the $\rm TEM_{00}$ mode is pumped externally, the interaction term can be linearized as
\be
\hat{H}_{\rm I}\approx\hbar G_0\hat{x}(\bar{a}^{*}\hat{b}+\bar{a}\,\hat{b}^{\dag}),
\ee
which is the same interaction considered in Ref. \cite{Marquart,Genes} after linearizing. Therefore, this three-mode
Hamiltonian can be mapped into an effective two-mode Hamiltonian by replacing $\hat{a}$ with its classical amplitude
$\bar{a}$. All the conclusions reached in the two-mode case are also valid here. The only difference is that the
carrier light TEM$_{00}$ is also on resonance rather than far detuned in the two-mode case \cite{Marquart,Genes}.
Therefore, we can achieve the resolved-sideband limit without compromising intra--cavity optical power. What's more,
the resonant condition $\omega_1-\omega_0=\omega_m$ here is equivalent to the frequency detuning equal to $\omega_m$ in the two-mode case.
This optimizes the energy transfer from the acoustic mode to the optical fields in the resolved-sideband limit \cite{Marquart,Genes}.

\section{Conclusion}

We have shown that a table top experiment based on a coupled cavity can be set up to realize self-sustained
three-mode opto-acoustic parametric interactions. Small adjustments of the positions of the optical
components enable us to continuously tune between instability and cooling regimes. This relative simple scheme
can be applied to design opto-acoustic amplifiers and experimentally investigate the three-mode parametric instability, a
possible issue of next-generation gravitational-wave detectors with high optical-power cavities. Besides,
it can also be used to realize the resolved-sideband cooling of acoustic modes with three-mode interactions.
\begin{center}
{\bf ACKNOWLEDGEMENTS}
\end{center}

We thank all the participants attending the Parametric Instability Workshop held at the Australian International
Gravitational Observatory (AIGO) site, especially Prof. S. P. Vyatchanin and Dr. S. E. Strigin for stimulating discussions
and giving precious advices on the improvement of the draft. We thank D. Price for pointing out several errors in the manuscripts.
Miao thanks Prof. Y. Chen for the invitation to visit the Albert-Einstein-Institut. The visit was supported by Alexander von
Humboldt Foundation's Sofja Kovalevskaja Programme. This research has been supported by the Australian Research Council
and the Department of Education, Science and Training and by the U.S. National Science Foundation. We thank the LIGO
Scientific Collaboration International Advisory Committee of the Gingin High Optical Power Facility for their supports.


\begin{thebibliography}{99}

\bibitem{Blair} D. G. Blair, E. N. Ivanov, M. E. Tobar,
P. J. Turner, F. van Kann and I.S. Heng, Phys. Rev. Lett. {\bf 74}, 1908 (1995);
\bibitem{Naik} A. Naik, O. Buu, M. D. LaHaye, A. D. Armour, A. A. Clerk, M. P. Blencowe and K. C. Schwab,
Nature {\bf 443}, 14 (2006);
\bibitem{Gigan} S. Gigan, H. R. B\"{o}hm, M. Paternostro, F. Blaser, G. Langer, J. B. Hertzberg, K. C. Schwab, D. B\"{a}uerle, M. Aspelmeyer and A. Zeilinger, Nature
{\bf 444}, 67 (2006);
\bibitem{Arcizet} O. ARcizet, P. F. Cohandon, T. Briant, M. Pinard and A. Heidmann, Nature {\bf 444}, 71 (2006);
\bibitem{Kleckner} D. Kleckner and D. Bouwmeester, Nature {\bf 444}, 75 (2006);
\bibitem{Schliesser1} A. Schliesser, P. DelHaye, N. Nooshi, K. J. Vahala, and
T. J. Kippenberg, Phys. Rev. Lett. {\bf 97}, 243905 (2006);
\bibitem{Corbitt1}  T. Corbitt, Y. Chen, E. Innerhofer, H. Muller-Ebhardt, D. Ottaway, H. Rehbein, D. Sigg, S. Whitcomb, C. Wipf and N. Mavalvala, Phys. Rev. Lett.
{\bf 98}, 150802 (2007);
\bibitem{Corbitt2} T. Corbitt, C. Wipf, T. Bodiya, D. Ottaway, D. Sigg, N. Smith,
S. Whitcomb, and N. Mavalvala, Phys. Rev. Lett. {\bf 99}, 160801 (2007);
\bibitem{Schliesser2} A. Schliesser, R. Rivi¨¨re, G. Anetsberger, O. Arcizet, T. J. Kippenberg. arXiv: 0709.4036v1 [quant-ph] (2007);
\bibitem{Poggio} M. Poggio, C. L. Degen, H. J. Mamin, and D. Rugar, Phys. Rev. Lett. {\bf 99}, 017201 (2007);
\bibitem{Thompson} J. D. Thompson, B. M. Zwickl, A. M. Jayich, F. Marquardt, S. M. Girvin and J. G. E. Harris, Nature {\bf 452}, 72 (2008);
\bibitem{Lowry} C. M. Mow-Lowry, A. J. Mullavey, S. Go$\beta$ler, M. B. Gray and D. E. McClelland, Phys. Rev. Lett. {\bf 100}, 010801 (2008);
\bibitem{Schediwy}  S. W. Schediwy, C. Zhao, L. Ju, D. G. Blair and P. Willems, Phys. Rev. A {\bf 77}, 013813 (2008);
\bibitem{braginsky} V. B. Braginsky, S. E. Strigin and S. P. Vyatchanin, Phys. Lett. A {\bf 287}, 331 (2001);
\bibitem{braginskyII} V. B. Braginsky, S. E. Strigin and S. P. Vyatchanin, Phys. Lett. A {\bf 305}, 111 (2002);
\bibitem{zhao} C. Zhao, L. Ju, J. Degallaix, S. Gras and D. G. Blair, Phys. Rev. Lett. {\bf 94}, 121102 (2005);
\bibitem{Ju} L. Ju, S. Gras, C. Zhao, J. Degallaix and D. G. Blair, Phys. Lett. A {\bf 354}, 360 (2006);
\bibitem{Gurkovsky} A. G. Gurkovsky, S. E. Strigin, S. P. Vyatchanin, Phys. Lett. A {\bf 362}, 91 (2007);
\bibitem{braginskyIII} V. B. Braginsky and  S. P. Vyatchanin, Phys. Lett. A {\bf 293}, 228 (2002);
\bibitem{Gras} S. Gras, D. G. Blair and L. Ju, Phys. Lett. A {\bf 372}, 1348 (2008);
\bibitem{AIGO}  C. Zhao, L. Ju, Y. Fan, S. Gras, B. J. J. Slagmolen, H. Miao, P. Barriga, D. G. Blair, D. J. Hosken, A. F. Brooks, P. J. Veitch, D. Mudge, and J.
    Munch, Phys. Rev. A {\bf 78}, 023807 (2008);
\bibitem{Guido} G. Mueller, LIGO document: G070441-00-R;
\bibitem{Marquart} F. Marquardt, J. P. Chen, A. A. Clerk, and S. M. Girvin, Phys. Rev. Lett. {\bf 99}, 093902 (2007);
\bibitem{Rae} I. Wilson-Rae, N. Nooshi, W. Zwerger, and T. J. Kippenberg, Phys. Rev. Lett. {\bf 99}, 093901 (2007);
\bibitem{Meers} B. J. Meers, Phys. Rev. D {\bf 38}, 8 (1988);
\bibitem{chen} A. Buonanno and Y. Chen, Phys. Rev. D {\bf 64}, 042006 (2001);
\bibitem{Rakhmanov} M. Rakhmanov, \textit{Dynamics of Laser Interferometric Gravitational Wave Detectors} (2000) (PhD Thesis);
\bibitem{Genes} C. Genes, D. Vitali, P. Tombesi, S. Gigan and M. Aspelmeyer, Phys. Rev. A {\bf 77}, 033804 (2008).

\end{thebibliography}
\end{document}